\documentstyle[12pt,aaspp4,epsf]{article}

\slugcomment{Submitted to The Astronomical Journal}

\lefthead{D. Merritt}
\righthead{Recovering Velocity Distributions}

\begin{document}

\title{Recovering Velocity Distributions via Penalized Likelihood}

\author{D. Merritt}
\affil{Department of Physics and Astronomy, Rutgers University,
    New Brunswick, NJ 08855}

\bigskip
\centerline{Rutgers Astrophysics Preprint Series No. 191}
\bigskip

\begin{abstract}
Line-of-sight velocity distributions are crucial for 
unravelling the dynamics of hot stellar systems.
We present a new formalism based on penalized likelihood for deriving 
such distributions from kinematical data, and evaluate the 
performance of two algorithms that extract $N(V)$ from 
absorption-line spectra and from sets of individual velocities.
Both algorithms are superior to existing ones in that the 
solutions are nearly unbiased even when the data are so poor that 
a great deal of smoothing is required.
In addition, the discrete-velocity algorithm is able to remove
a known distribution of measurement errors from the estimate of 
$N(V)$.
The formalism is used to recover the velocity distribution of
stars in five fields near the center of the globular cluster 
$\omega$ Centauri.
\end{abstract}

\section{Introduction}

The most complete kinematical information obtainable for a 
distant stellar system is the distribution of line-of-sight 
velocities at every point in the image.
Velocity distributions are crucial for understanding the 
dynamical states of slowly-rotating stellar systems like 
elliptical galaxies and 
globular clusters, since velocity dispersions alone 
place almost no constraints on the form of the potential 
unless one is willing to make ad hoc assumptions about the shape 
of the velocity ellipsoid (\cite{dem92}).
Velocity distributions are also useful when searching for kinematically 
distinct subcomponents (e.g. \cite{fri88}; \cite{rif92}). 

The velocity distribution at point ${\bf R}$ in the image of a 
stellar system, $N({\bf R},V)$, can be related to the 
data in different ways depending on the nature of the 
observations.
In a system like a globular cluster, for which the data usually consist 
of individual stellar velocities, the velocity distribution is 
just the frequency function of stellar velocities defined by 
those stars with apparent positions near to ${\bf R}$.
Since measured velocities are always in error, the observed and 
true $N(V)$'s are related via a convolution integral.
In a distant, unresolved galaxy, one typically measures the integrated 
spectrum of many stars along a line of sight.
The observed spectrum is then a convolution of the velocity 
distribution of these stars with the broadening function of the 
spectrograph, and the spectrum of a typical star.

With both sorts of data, the goal is to find a function $N(V)$, at 
some set of points ${\bf R}$, such that $\sum_i L\{Y_i;N(V)\}$ -- 
the log likelihood of observing the data $Y_i$ given 
$N$ -- is large.
Maximizing this quantity over the space of all possible 
functions $N(V)$ is unlikely to yield useful results, however, 
since any $N(V)$ that maximizes the likelihood (assuming it exists, which
it often will not) is almost certain to be extremely noisy.
This is obviously true if the data are related to the model via
a convolution, since the process of deconvolution will amplify the 
errors in the data.
But it is equally true if $N(V)$ is simply the frequency function of 
observed velocities, since the most likely distribution 
corresponding to an observed set of $V$'s is just a sum of delta 
functions at each of the measured velocities.
One is therefore forced to place smoothness constraints on the 
solution.

But smoothing always introduces a bias, i.e. a systematic deviation 
of the solution from the true $N(V)$.
The nature of the bias is obvious when the smoothing is carried 
out by imposing a rigid functional form on $N(V)$, since 
the true function will almost certainly be different from this 
assumed form.
But even nonparametric smoothing generates a bias since it 
effectively averages the data over some region.
Furthermore, because the required degree of smoothing increases with the 
amplitude of the noise in the data, the error from the bias goes 
up as the quality of the data falls.
An ideal algorithm for estimating $N(V)$ would therefore be one 
in which the bias introduced by the smoothing was effectively minimized, 
so that the derived $N(V)$ was close to the true function even when the data 
were so poor that a great deal of smoothing was required.

One way to accomplish this is to make use of prior knowledge 
about the likely form of $N(V)$.
Many studies of stellar and galactic systems have shown that 
$N(V)$ is often close to a Gaussian.
This fact suggests that we infer $N(V)$ by maximizing a quantity 
like
\begin{equation}
\log{\cal L}_p = \sum_i L\{Y_i;N(V)\} - \alpha P(N) ,
\end{equation}
the ``penalized log likelihood,''
where the penalty functional $P(N)$ is large for any $N(V)$ that 
is noisy and zero for any $N(V)$ that is Gaussian.
A natural choice for such a penalty functional has been suggested 
by \cite{sil82}:
\begin{equation}
P_G(N) = \int_{-\infty}^{+\infty}\left[(d/dV)^3\log N(V)\right]^2 dV. 
\end{equation}
This functional assigns zero penalty to any $N(V)=N_0 
\exp[-(V-V_0)^2/2\sigma^2]$, i.e. any Gaussian velocity distribution,
and a large penalty to any $N(V)$ that is rapidly varying.
The limiting estimate as the smoothing parameter $\alpha$ tends 
to infinity is the normal distribution that best corresponds, in a 
maximum-likelihood sense, to the data.
Thus varying $\alpha$ takes one from an estimate of $N(V)$ that 
is very noisy but that reproduces the data well, to the ``infinitely 
smooth'' maximum likelihood Gaussian fit to $N(V)$.
When the data are copious and accurate, the degree of smoothing 
required, i.e. the value of $\alpha$, will be small and the inferred 
$N(V)$ will be close to the true function.
When the data are poor, $\alpha$ must be increased to deal with the noise; 
however even a 
very large value of $\alpha$ will yield an $N(V)$ that is 
nearly Gaussian and that is therefore likely to be close to the true velocity 
distribution.
Regardless of the quality of the data, there will always be a formally 
optimal choice of $\alpha$ that yields a solution that is 
closest in some sense to the true $N(V)$ -- neither too noisy 
nor too biased.

Here we apply penalized likelihood methods 
to the recovery of velocity distributions from kinematical data 
of two sorts: Doppler-broadened spectra, and discrete velocities.
There are of course a large number of excellent algorithms 
already in use for extracting velocity distributions from 
absorption-line spectra 
(\cite{fri88}; \cite{ben90}; \cite{riw92}; \cite{kum93}; 
\cite{vdf93}; \cite{saw94}).
Is another approach really needed?
Many of the existing algorithms are essentially nonparametric and 
hence yield unbiased estimates of $N(V)$ when the 
signal-to-noise ratio of the data is large.
Their behavior given data with small S/N is more variable, 
however, due to the different ways in which they carry out the 
smoothing.
For instance, in algorithms that represent the unknown $N(V)$ via 
a basis-set expansion,
\begin{equation}
N(V) = \sum_{k=1}^K C_k N_k(V),
\end{equation}
the smoothing is accomplished by truncating the expansion after 
a finite number of terms.
Adding more terms decreases the bias by giving the algorithm 
more freedom to match the true $N(V)$;
however the level of fluctuations in the solution increases as the bias
falls since the high-order terms will always try to reproduce the noise 
in the data.
The optimal estimate for data of a given quality will therefore 
contain only a limited number 
of terms, and if the data are poor, this number will be small; 
hence the solution will be biased toward the functional form 
selected for $N_1(V)$.
Of course one can choose $N_1$ to be the normal distribution
(e.g. \cite{vdf93}) but its mean and variance
must be specified, and the optimal choices for both parameters will 
depend in some complex way on the number of terms retained 
in the expansion and on the level of noise in the data.
Worse, one does not have complete freedom in this approach to 
adjust the smoothing to its optimal level since the basis 
set is fixed and its terms are discrete.

Most of the spectral deconvolution schemes currently in use have
continuously-adjustable smoothing and so do not suffer from 
this latter defect.
However few of these schemes incorporate any prior 
knowledge about the likely form of $N(V)$ and so the solutions 
they return given noisy data tend to be unphysical, 
with functional forms that are determined primarily by the 
mechanics of the smoothing.
(In fact it is a common practice to fall back on parametric 
methods when the data are poor.)
For instance, both the Wiener-filtered algorithm of Rix \& White 
(1992) and the kernel-based algorithm of Kuijken \& Merrifield (1993) 
yield $N(V)\approx {\rm constant}$ in the limit of infinite smoothing.
Since real velocity distributions are unlikely to be well approximated
by constant functions, the estimates produced by these schemes 
can be substantially biased.

The approach advocated here is neither better nor worse than the 
existing ones when the data are of high quality.
However it has an advantage when the data are poor, 
since even a large degree of smoothing is likely to bias
the solution only slightly.
Furthermore the amount of smoothing can be adjusted with 
infinite precision by varying $\alpha$.
In practice it may be difficult to find the optimal choice of 
$\alpha$ given noisy data, but there are bootstrap
techniques for choosing $\alpha$ that work well in most cases; 
and one can always elect to be guided by physical intuition 
when deciding how smooth $N(V)$ should be.
Most important, even a gross overestimate of $\alpha$ will yield 
no worse than a Gaussian fit to the velocity distribution, which
is perhaps the best that can be hoped for when the signal-to-noise
ratio or the number of observed velocities is small.

(Much of the uncertainty in $N(V)$'s derived from absorption line 
spectra is due to systematic errors such as incorrect 
continuum subtraction, template mismatch, limited resolution, etc.
These problems afflict every spectral deconvolution scheme 
and have been treated at length by other authors.
We have nothing new to say about these systematic effects and will 
simply ignore them -- our focus is on the uncertainty in $N(V)$ that 
is generated by noise in the data rather than by systematic 
errors.)

Below we evaluate the performance of penalized-likelihood algorithms 
for estimating $N(V)$ from spectral data (\S2) and from 
discrete velocities (\S3).
The former sort of data are routinely obtained for distant galaxies and 
the latter for globular clusters, clusters of galaxies and 
systems of emission-line objects around galaxies.
The same formalism works equally well in both cases; essentially
all that needs to be changed is the form of the matrix that relates the 
observable quantities to $N(V)$.
We also show how the cross-validation score can be used to estimate 
the optimal degree of smoothing from the data.
We then (\S4) apply the formalism to the recovery of the velocity 
distribution near the center of the globular cluster $\omega$ Centauri 
using a new sample of stellar radial velocities from an imaging 
spectrophotometer.

\section{Spectra}

The optical spectrum at any point in the image of a galaxy is 
made up of the sum of the spectra of its component stars along 
the line of sight, Doppler shifted according to their line-of-sight 
velocities $V$.
If the galaxy spectrum is $I(\lambda)$ and the spectrum of a 
single star -- after convolution with the instrumental response -- 
is $T(\lambda)$, then
\begin{equation}
I(\ln\lambda)=\int_{-\infty}^{+\infty} N(V) T\left(\ln\lambda-V/c\right)
\ dV = N\circ T,
\end{equation}
the convolution of $N(V)$ with the template.
We accordingly seek a function $N = \hat N(V)$ such that
\begin{equation}
-\log{\cal L}_p = \sum_i [I(\lambda_i) - (N\circ T)_i]^2 + \alpha P_G(N)
\label{penlik}
\end{equation}
is minimized, where $I(\lambda_i)$ is the observed spectral intensity at 
wavelength $\lambda_i$.

We first define a discrete grid in velocity space, $V_1,...,V_m$, where 
$V_j=V_1+(j-1)\delta$ and $\delta=(V_m-V_1)/(m-1)$.
Typically $V_1\approx \overline{V} - 4\sigma$ and $V_1\approx 
\overline{V} + 4\sigma$, where $\overline{V}$ is the mean 
velocity of the system and $\sigma$ is the expected 
velocity dispersion.
The number of grid points $m$ should be large enough that the 
numerical solution to the minimization problem is essentially 
independent of $m$, yet small enough that the minimization --- 
which requires a computational time that scales as 
$\sim m^3$ --- does not take inordinately long.
We mostly used $m=50$ velocity grid points in what follows.
A discrete representation of the right hand side of 
Eq. (\ref{penlik}) on this grid is
\begin{equation}
\sum_i [I(\lambda_i) - (N\circ T)_i]^2 +  \alpha\delta^{-5}
\sum_{j=2}^{m-2}\left[ -\log N_{j-1}+3\log N_j-3\log N_{j+1}+\log 
N_{j+2}\right]^2.
\label{sildisc}
\end{equation}
In the first term, $N\circ T$ is computed by 
fitting an interpolating spline to the template spectrum and 
assuming a linear dependence of the (unknown) function $N$ on $V$ 
between the grid points.
The convolution can then be represented as a matrix 
multiplication, i.e.
\begin{equation}
(N\circ T)_i \approx \sum_j A_{ij}N_j,
\label{matrix}
\end{equation}
\begin{equation}
A_{ij} = {V_{j+1}I_{ij} - J_{ij}\over V_{j+1}-V_j} - 
{V_{j-1}I_{i,j-1} - J_{i,j-1}\over V_j - V_{j-1}},
\end{equation}
\begin{equation}
I_{ij} = \int_{V_j}^{V_{j+1}} T\left(\ln\lambda_i - V/c\right) 
dV,
\end{equation}
\begin{equation}
J_{ij} = \int_{V_j}^{V_{j+1}} V\ T\left(\ln\lambda_i - V/c\right) 
dV.
\end{equation}
One then varies the $m$ parameters $N_j$ until a
minimum is obtained; the use of $\log N$ in the penalty 
functional guarantees that the solution will remain everywhere 
positive without the need to impose additional positivity constraints.
Minimization was carried out using the NAG routine E04JAF and 
required about one minute on a DEC Alpha 3000/700 machine.

If the true $N(V)$ is a Gaussian, the optimal estimates will be 
obtained from this algorithm by choosing $\alpha$ to be very large -- 
the algorithm becomes essentially parametric in this limit, 
returning the normal distribution that best approximates the 
data.
But we would like to show that the algorithm works well even in cases 
where $N(V)$ deviates strongly from a Gaussian.
The crucial problem, of course -- both for this algorithm and any 
other -- lies in choosing the correct degree of smoothing.

Figure 1 shows three estimates of $N(V)$ based on pseudo-data 
generated from the very non-Gaussian velocity distribution
\begin{equation}
N(V) = {1\over\sqrt{2\pi}\sigma}\left(0.7\exp[{-(V-V_1)^2/2\sigma^2}] + 
0.3\exp[{-(V-V_2)^2/2\sigma^2}]\right),
\label{mkbf}
\end{equation} 
with $V_1=140$km s$^{-1}$, $V_2=-140$km s$^{-1}$ and $\sigma=80$ km 
s$^{-1}$.
This $N(V)$ was found by \cite{mek94} to 
provide a good description of the stars in the disk of the Sab galaxy 
NGC 7217.
Merrifield \& Kuijken's broadening function was convolved 
with a template spectrum 
from a K0III star, kindly provided by T. Williams; to this broadened 
spectrum was added Gaussian random noise with 
amplitude $\sigma_N$ such that $I/\sigma_N = {\rm S/N} = 20$.
The resultant spectrum was then sampled at 800 points in the wavelength
range $4800\AA<\lambda<5500\AA$ and these ``data'' were given to 
the minimization routine.
Figure 1 illustrates the famous ``bias-variance'' tradeoff of 
solutions to ill-conditioned problems.
When $\alpha$ is small, the estimated $N(V)$ matches the true 
function in an average sense, but fluctuates strongly from grid point to 
grid point.
For larger $\alpha$, the estimate is nearly correct, with 
fluctuations of only a few percent around the true $N(V)$.
When $\alpha$ is increased still more, the solution is driven 
toward a single Gaussian with roughly the same mean and 
dispersion as that of the true $N(V)$.
All of these estimates for $N(V)$ reproduce the input spectrum 
with acceptable accuracy, but $N(V)$ itself is correctly 
recovered only when $\alpha$ is chosen appropriately -- 
a characteristic feature of ill-conditioned problems.
Nevertheless an appropriate choice of $\alpha$ allows one to 
recover even this very non-Gaussian $N(V)$ with good accuracy.

Figure 1 also shows three attempts to recover the broadening function
\begin{equation}
N(V) = {1\over\sqrt{2\pi}}\left[{0.4\over\sigma_1}\exp\left({-V^2/2\sigma_1^2}
\right) + 
{0.6\over\sigma_2}\exp\left({-V^2/2\sigma_2^2}\right)\right]
\label{peak}
\end{equation} 
with $\sigma_1=200$ km $s^{-1}$, $\sigma_2=50$ km s$^{-1}$.
This $N(V)$ is more peaked than a Gaussian, and is 
designed to represent the velocity distribution in the halo of a 
spherical galaxy dominated by radial orbits (\cite{dej87}).
Once again the nature of the solution depends on $\alpha$, with 
large values of $\alpha$ generating a Gaussian approximation to 
the true broadening function.
But the optimal $\alpha$ produces an estimate of $N(V)$ that is 
again quite close to the true function, and the velocity dispersion 
derived from the 
estimated $N(V)$ remains close to the true dispersion even when 
$\alpha$ is far from its optimal value.

The examples just discussed were based on data with a high 
signal-to-noise ratio.
A more stringent test is the recovery of a non-Gaussian $N(V)$ 
from noisy data.
Since a larger degree of smoothing will be required, the penalty 
functional will tend to drive the solution away from the true 
$N(V)$ as the level of the noise increases.
Figure 2 shows how the average error in the estimated $N(V)$ 
varies with the signal-to-noise ratio of the data.
For each value of S/N, 300 random realizations of the same spectrum 
were generated from the Merrifield-Kuijken broadening function
and the value of $\alpha$ that minimized the 
average, integrated square deviation between the estimates $\hat N$ and 
the true function $N$ was found.
The integrated square error was defined as 
\begin{equation}
{\rm ISE} = {\int \left[N(V) - \hat N(V)\right]^2 dV\over\int N^2(V) 
dV};
\label{ISE}
\end{equation}
the normalizing function in the denominator was added so that a 
value ISE $\approx 1$ corresponds to an rms deviation of 
roughly 100\% between $N$ and $\hat N$.
The MISE was then defined as the mean value of the ISE over the 
300 realizations using this 
``optimal'' value of $\alpha$.
Figure 2 shows that the MISE falls roughly as a power law with S/N, 
with a logarithmic slope of approximately $-1.3$.
The same information is presented in more detail in Figure 3, 
which shows the average estimate and its 95\% variability 
bands for each value of S/N.
This figure suggests that the existence of two peaks in $N(V)$ is 
recoverable for S/N $> 5$.

In these examples, the optimal choice of $\alpha$ and its 
associated MISE could be computed since the correct form of 
$N(V)$ was known.
In general one does not know $N(V)$, of course, and the choice of 
$\alpha$ must be based somehow on information 
contained within the data itself.
One way of doing this is via the ``cross-validation score'' 
(\cite[p. 30]{grs94}; \cite[p. 47]{wah90}).
One seeks the value of $\alpha$ that minimizes
\begin{equation}
{\rm CV}(\alpha) = {1\over n}\sum_{i}\left[I(\lambda_i) - (N^{[i]}\circ 
T)_i\right]^2,
\end{equation}
where $N^{[i]}$ is the minimizer of Eq.
(\ref{sildisc}) with the $i$th data point left out.
In effect, the CV measures the degree to which the spectral 
intensities predicted by $N(V)$ are consistent with the observed intensities.
This is not quite the same as asking how close the estimated 
$N(V)$ is to the true $N(V)$, but it is probably the best that 
one can do in practice (\cite{wah80}).

Figure 4 shows the result of two attempts to recover the optimal $\alpha$ from 
fake spectra generated using the Merrifield-Kuijken broadening 
function with S/N$=5$ and $10$.
The values of $\alpha$ that minimize the CV are tolerably close 
to the values that actually minimize the ISE.
More to the point, the ISE of the $N(V)$'s generated using the CV 
estimates of $\alpha$ differ only negligibly from those obtained 
using the optimal $\alpha$'s.
These examples suggest that one can indeed hope to recover a useful 
estimate of the optimal smoothing parameter from the data alone.
At the very least, such an estimate would provide a starting 
point when searching for the $\alpha$ the produces the 
physically most appealing estimate of $N(V)$.
  
\section{Discrete Velocities}

In many stellar systems, information about $N(V$) is most naturally 
obtained in the form of discrete velocities.
If the velocities are measured with negligible error, $N(V)$ is 
simply their frequency function, which can be defined as the 
function that maximizes the penalized log-likelihood
\begin{equation}
\log{\cal L}_p = \sum_{i=1}^n \log N(V_i) - \alpha P_G(N)
\end{equation}
subject to the constraints
\begin{equation}
\int N(V) dV = 1, \ \ \ \ N(V) \ge 0
\end{equation}
(\cite[p. 102]{tht90}).
The penalty functional is needed since, in its absence, the optimal 
estimate would be a set of delta-functions at the measured velocities.

But the uncertainty in the measured velocities is sometimes comparable 
to the width of $N(V)$.
For instance, radial velocities of faint stars in globular clusters 
may have measurement errors of a few km s$^{-1}$ 
compared to intrinsic velocity dispersions of $\sim 10$ km s$^{-1}$.
The observable function is then not $N(V)$ but rather its convolution with
the error distribution, $N\circ P$.
Assuming that the errors have a normal distribution, with dispersion 
$\sigma_N$, we have
\begin{equation}
N\circ P = {1\over\sqrt{2\pi}\sigma_N}\int_{-\infty}^{+\infty} N(V') 
\exp\left[-(V-V')^2/2\sigma^2\right]\ dV'
\end{equation}
and one accordingly seeks the $N(V)$ that maximizes
\begin{equation}
\log{\cal L}_p = \sum_{i=1}^n \log (N\circ P)_i - \alpha P_G(N)
\end{equation}
subject to the same constraints on $N$.
Following Silverman (1982), we find the solution to this 
constrained optimization problem as the unconstrained maximizer of 
the functional
\begin{equation}
\sum_{i=1}^n \log (N\circ P)_i - \alpha P_G(N) - 
n\int N(V)dV.
\label{sildisc2}
\end{equation}
This problem is formally very similar to the deconvolution 
problem solved above, 
and a discrete representation of (\ref{sildisc2}) on a grid in 
velocity space is
\begin{equation}
\sum_{i=1}^n \log (N\circ P)_i +  \alpha\delta^{-5}
\sum_{j=2}^{m-2}\left[ -\log N_{j-1}+3\log N_j-3\log N_{j+1}+\log 
N_{j+2}\right]^2 - n\sum_{j=1}^m \epsilon_jN_j,
\end{equation}
with $\epsilon_1=\epsilon_m=\delta/2$, $\epsilon_j=\delta,j=2,...,m-1$.
The convolution of $N$ with $P$ is again represented as a matrix,
Eq. (\ref{matrix}), with
\begin{equation}
I_{ij} = {1\over\sqrt{2\pi}\sigma_N}
\int_{V_j}^{V_{j+1}} \exp\left[{-(V-V_j)^2\over 
2\sigma_N^2}\right] dV,
\end{equation}
\begin{equation}
J_{ij} = {1\over\sqrt{2\pi}\sigma_N}\int_{V_j}^{V_{j+1}}V\ 
\exp\left[{-(V-V_j)^2\over 2\sigma_N^2}\right] dV,
\end{equation}
which can be expressed in terms of the error function.

Noise in the data now comes from two sources: measurement errors, 
as described by $\sigma_N$; and finite-sample fluctuations due to 
the limited number $n$ of measured velocities.
Figure 5 shows how the MISE depends on $n$ for
data sets generated from the velocity distribution
\begin{equation}
N(V) = {0.5\over\sqrt{2\pi}\sigma}\left(\exp[{-(V-V_1)^2/2\sigma^2}] + 
\exp[{-(V-V_2)^2/2\sigma^2}]\right),
\label{flat}
\end{equation} 
with $V_1=9$ km s$^{-1}$, $V_2=-9$ km s$^{-1}$ and $\sigma=8$ km 
s$^{-1}$.
This flat-topped velocity distribution was designed to mimic 
$N(V)$ near the projected center of a globular cluster
containing an abundance of nearly-radial orbits.
(Examples of such models and their velocity distributions 
may be found in \cite{mer89}).

The two sets of points in Figure 5 correspond to pseudo-data 
generated with zero velocity errors (circles) and with $\sigma_N=4$ km 
s$^{-1}$ (squares); the latter value is roughly one-third of the 
intrinsic velocity dispersion.
The MISE falls off roughly as $n^{-0.9}$ for both types of data,
almost as steep as the $n^{-1}$ dependence of parametric estimators. 
Of course the mean error is larger, at a given $\alpha$, for the sample
with nonzero $\sigma_N$;
however an increase in sample size from $n=200$ to $n=300$ produces
the same decrease in the MISE as a reduction in the measurement 
uncertainties from $4$ to $0$ km s$^{-1}$.
Thus even relatively large measurement errors can 
be overcome by a modest increase in sample size (assuming, of
course, that the distribution of errors is well understood).
Figure 6 shows the average estimates obtained with the 
optimal smoothing parameters and their 95\% variability bands.
The non-Gaussian nature of $N(V)$ is surprisingly well 
reproduced even for $n=200$, but the two peaks only begin to 
be clearly resolved for $n=1000$.

Various techniques, including a version of the cross-validation 
score described above, can be used to estimate optimal smoothing 
parameters for data like these.
The ``unbiased'' or ``least-squares'' cross-validation score
(Scott 1992, p. 166; Silverman 1986, p. 48) is defined as
\begin{equation}
{\rm UCV}(\alpha) = \int(\hat N\circ P)^2 dV - {2\over n}\sum_{i=1}^n 
(\hat N^{[i]}\circ P)_i
\label{UCV}
\end{equation}
where $\hat N^{[i]}$ is an estimate of $N$ 
obtained by omitting the $i$th velocity.
The value of $\alpha$ that minimizes the UCV is an estimate of the 
value that minimizes the ISE of $\hat{N}\circ P$.
Figure 7 shows the dependence of the UCV on $\alpha$ for two data 
sets, with $n=200$ and $n=500$, generated from the velocity 
distribution of Eq. (\ref{flat}) with $\sigma_N=0$.
The minimum in both curves occurs at a value of $\alpha$ close to 
the value that actually minimizes the ISE.

The results just presented suggest that increasing the number of 
measured velocities in 
globular clusters may be a greater priority than reducing measurement errors 
if the goal is to determine $N(V)$, since existing techniques can already 
extract stellar radial velocities with greater precision
than the uncertainty $\sigma_N=4$ km s$^{-1}$ adopted here.
Because one would like to estimate $N(V)$ at several different 
points in an image, the total number of velocities 
required for a single globular cluster would be in the thousands at least.

\section{Application to $\omega$ Centauri} 

Fortunately, data sets of this size are now becoming available for a 
number of stellar systems.
Here we analyze radial velocities of a new sample of 4200 stars near 
the center of the globular cluster $\omega$ Centauri.
The velocities were measured using the Rutgers Fabry-Perot 
interferometer on the CTIO 1.5m telescope and were kindly made 
available by C. Pryor, who also carried out an analysis of
the measurement errors.
Figure 8 shows the spatial distribution of the observed stars.
The effective field of view of the Fabry-Perot is about 2.75 arc minutes 
in radius
and is offset by about 1.1 arc minutes E/SE from the cluster center as 
determined by Meylan et al. (1995).
The core radius of $\omega$ Centauri is around 
2.5 arc minutes (\cite{pek75}) so the observed 
velocities lie mostly within the projected core.

The core of a globular cluster is perhaps not a very auspicious
place to look for non-Gaussian velocity distributions.
However $\omega$ Centauri is fairly young in a 
collisional-relaxation sense, 
with an estimated central relaxation time of a few billion years
(\cite{mem95}).
Thus its velocity distribution might still be non-Maxwellian.
Furthermore there is evidence for two, chemically distinct 
populations in $\omega$ Centauri (Norris et al. 1996) which may 
have formed at different epochs and hence may have different 
kinematics.
Finally, we note that the theoretical expectation of Maxwellian 
velocity distributions in globular clusters has rarely been 
tested by direct determination of line-of-sight velocity distributions.
For this reason alone, it seems worthwhile to estimate $N(V)$ in 
$\omega$ Centauri.

About 20 stars were removed from the original sample because 
their Fabry-Perot line profiles showed contamination by H$\alpha$ emission.
Velocity uncertainties of the remaining stars were estimated using 
standard procedures (e.g. \cite{gep94}); a typical estimated error was 
3-5 km s$^{-1}$.
Although the estimated errors were found to correlate with stellar magnitude,
this fact was ignored in the analysis and $\sigma_N$ was simply 
set equal to the rms estimated uncertainty of all the stars in each 
subsample, about 4 km s$^{-1}$ in each case.
For comparison, the central velocity dispersion of $\omega$ Centauri is
15-20 km s$^{-1}$ (Meylan et al. 1995).

One would like to estimate $N(V)$ independently at many different 
points in this observed field of view.
However Figure 6 suggests that sample sizes less than a few 
hundred are not very useful for detecting departures from normality.
The compromise, as illustrated in Figure 8, was to identify five, 
partially overlapping fields of one arc minute radius
containing about one-half of the observed stars.
Two fields lie along the estimated rotation axis of the cluster, 
at a position angle of $30^{\circ}$ east from north with their 
centers displaced one arc minute from the cluster center.
The three additional fields were situated along a perpendicular 
line.
The orientation of the rotation axis was estimated from 
this sample, but is consistent with a determination based on 
$\sim$ 500 stars with velocities measured by CORAVEL 
(\cite{mey96}).
The offset between the Fabry-Perot field of view and the cluster 
center was used to advantage by centering the fifth field at a 
distance of 2.5 arc minutes from the cluster center along the direction 
of maximal rotation.

The number of stars in each of the subsamples, along with their 
mean velocity and velocity dispersion (with estimated errors 
removed) in km s$^{-1}$, are given in Table 1.
The average number of stars per field was 653 for the four inner 
fields, with 456 stars in Field 5.

The velocity distributions in Fields 1 and 5 exhibit the greatest 
apparent departures from normality.
Figure 9 shows the dependence of the inferred $N(V)$ in these two 
fields on the value of the smoothing parameter $\alpha$.
Figure 10 displays estimates of $N(V)$ in all five fields, for one 
choice of $\alpha$, and their estimated 95\% confidence bands.
The confidence bands were computed in the usual way via the bootstrap
(Scott 1992, p. 259); the choice of $\alpha$ is justified below.
Also shown are the normal distributions with the same mean and 
variance as the inferred $N(V)$'s.

None of the recovered velocity distributions is strikingly
non-Gaussian, although the confidence bands in Fields 1 and 5 do 
barely exclude the normal distribution at one or more points.
The inferred $N(V)$ for Field 5 is more centrally peaked than a 
Gaussian and has what might be described as a tail or bump at large 
positive velocities.
It is tempting to interpret this curve as resulting from the 
superposition of two normal distributions with different means 
and variances, though such an interpretation seems extravagant 
given the relatively small amplitude of the deviations.

The choice of the optimal $\alpha$ for these estimates presented 
certain difficulties.
When dealing with random samples drawn from the normal 
distribution, the value of $\alpha$ that minimizes the error in
an estimate of $N(V)$ would clearly be very large, since an infinite 
$\alpha$ always returns a normal distribution.
Because the velocity distributions in $\omega$ Centauri do appear 
to be quite Gaussian, we might expect the optimal $\alpha$'s to be 
``nearly infinite'' and hence difficult to estimate from the data 
alone.
In fact the UCV score (Eq. \ref{UCV}) 
was found not to have a minimum at any value of $\alpha$ in any 
of these five sub-samples; instead the function ${\rm UCV}(\alpha)$ 
was always found to asymptote to a constant value as $\alpha$ was 
increased.
A similar result was obtained using the ``likelihood 
cross-validation'' score (Silverman 1986, p. 53).
These results certainly do not imply that the distribution of 
velocities in $\omega$ Centauri must be exactly Gaussian, since 
cross-validation is not a precise prescription for determining $\alpha$ 
and often fails to give an extremum.
But it appears that these velocity distributions are close enough
to Gaussian that the cross-validation technique can not find a 
significant difference between the estimates made with finite and 
infinite $\alpha$.

The following alternative scheme was adopted for selecting the 
optimal $\alpha$.
A crude estimate of $N\circ P$ can be obtained by replacing each 
of the measured velocities by a kernel function of fixed width.
If $N\circ P$ is exactly Gaussian, and if the kernel is also 
Gaussian, the optimal window width (i.e. dispersion) for the 
kernel may be shown to be 
\begin{equation}
h_{opt}=1.06\sigma n^{-1/5}
\end{equation}
(Silverman 1986, p. 45).
Fixed-kernel estimates of $N\circ P$ in each of the five fields 
were generated from the data using this optimal $h$, and compared 
to estimates of $N\circ P$ using the penalized-likelihood algorithm
with various values of $\alpha$.
These comparisons could not be made precise since the fixed-kernel 
estimates have a larger bias and do not compensate for the 
velocity measurement errors.
Nevertheless it is reasonable to assume that the degree of 
``roughness'' in the optimal kernel-based estimates ought to be 
similar to that in the penalized-likelihood estimates made with 
the optimal value of $\alpha$.
The plots of $N(V)$ in Fig. 10 were made using these estimates of 
the optimal $\alpha$'s.

We conclude from this analysis that the evidence for non-Gaussian velocity 
distributions in our five fields near the center of $\omega$ Centauri 
is marginal at best.
We note that the strongest deviations appear in the field that is 
farthest from the center.
Perhaps a sample of stars even farther from the core -- where the 
relaxation time exceeds the age of the universe -- would show even 
larger departures from normality.

\bigskip\bigskip
The $\omega$ Centauri radial velocities analyzed here were obtained
by K. Gebhardt, J. Hesser, C. Pryor and T. Williams using the Rutgers 
Fabry-Perot interferometer.
C. Pryor devoted considerable time to reducing the observations 
and estimating the measurement uncertainties in this 
sample.
Conversations with C. Joseph, M. Merrifield, H.-W. Rix, P. Saha, R. van der 
Marel and T. Williams were very helpful for understanding 
the ins and outs of spectral deconvolution.
C. Pryor read parts of the manuscript and made useful suggestions for 
improvements.
This work was supported by NSF grant AST 90-16515 and NASA grant 
NAG 5-2803.

\clearpage

\begin{table}
\caption{Fields in $\omega$ Centauri}
\medskip
\begin{tabular}{crcc} \hline
Field	&   N	&   $\langle V\rangle$	&   $\sigma$ \\ \hline
1	&   655 &	0.		&	16.0 \\
2	&   670 &	-2.7		&	16.7 \\
3	&   624 &	0.1		&	16.8 \\
4	&   663 &	2.2		&	16.6 \\
5	&   456 &	6.4		&	14.6 \\	\hline	
\end{tabular}
\end{table}

\clearpage

\figcaption[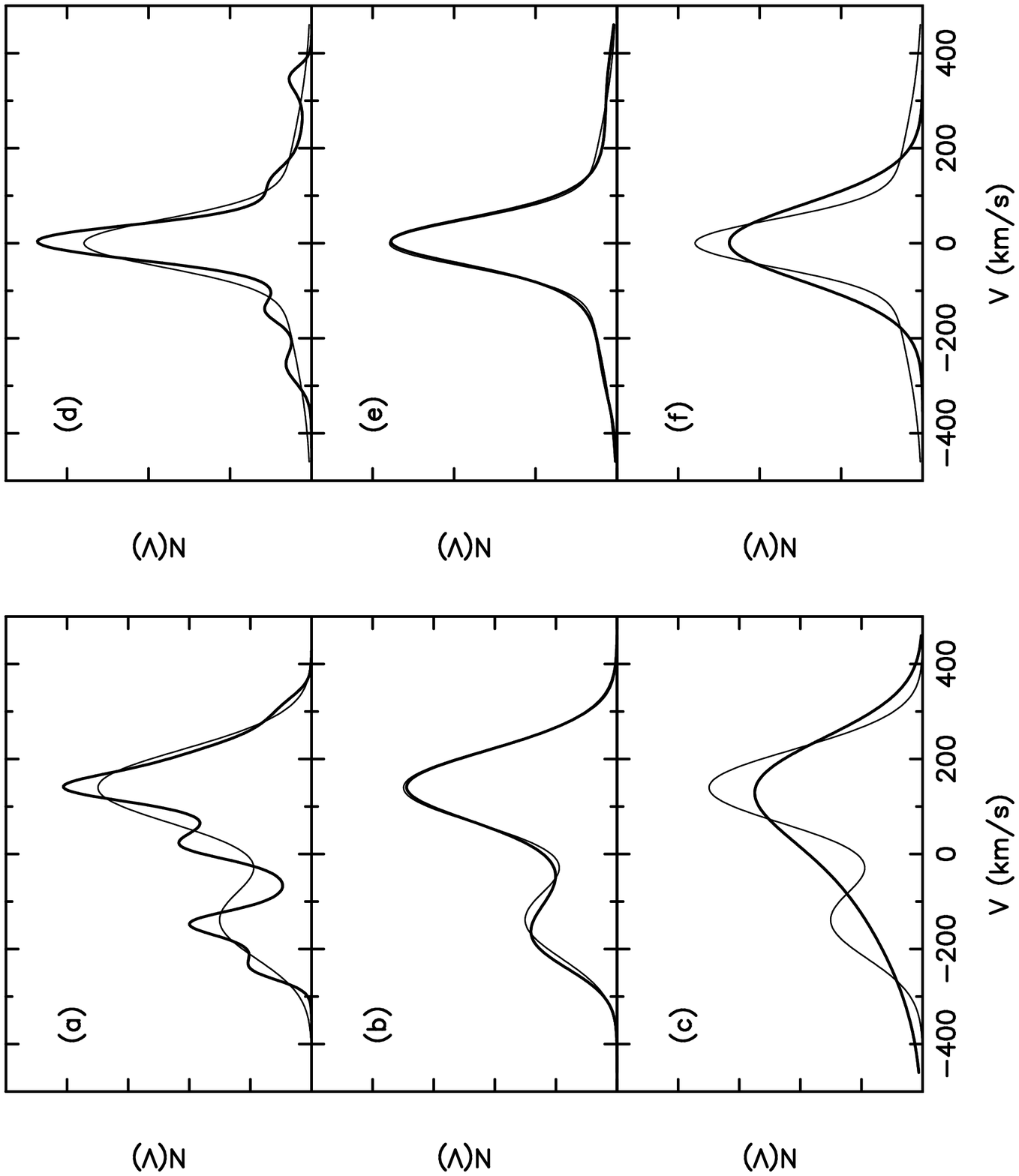]{\label{fig1}} Bias-variance tradeoff in
estimates of non-Gaussian $N(V)$'s from absorption line spectra. 
Input spectra were generated using two different broadening functions
(thin curves) with added noise of amplitude S/N=20.
(a) - (c): Merrifield-Kuijken broadening function, Eq. 
(\ref{mkbf});
(d) - (f): broadening function of Eq. (\ref{peak}).
$N(V)$ was then estimated (thick curves) 
using three different values of $\alpha$.
(a), (d): undersmoothed; (b), (e): optimally smoothed;
(c), (f): oversmoothed.
In the limit of large $\alpha$, i.e. infinite smoothing,
the estimates tend toward a
Gaussian with approximately the same mean and dispersion as
the true $N(V)$.

\figcaption[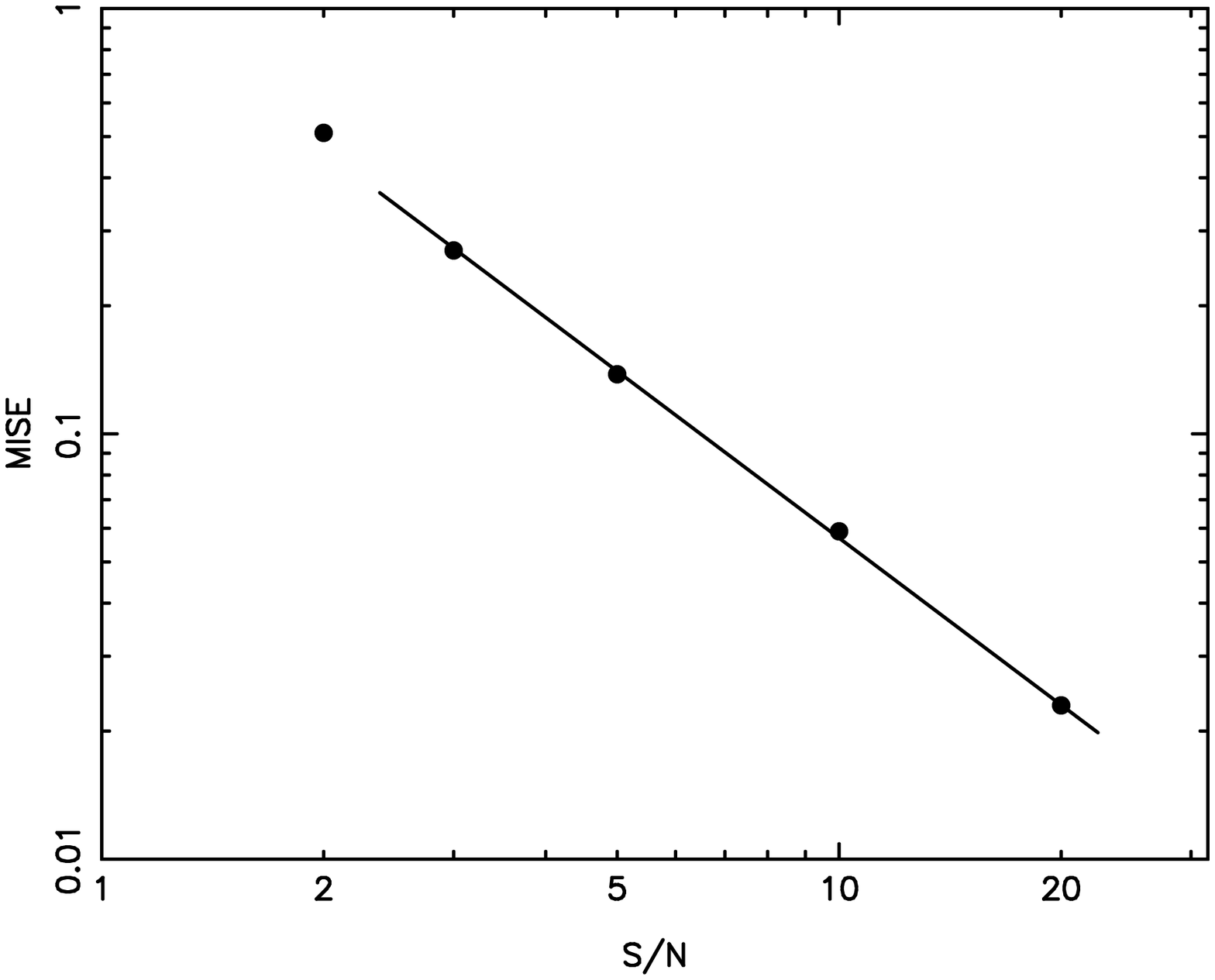]{\label{fig2}} Dependence of the 
mean integrated square error of 
the recovered broadening function on the signal-to-noise ratio of
the spectrum, for spectra generated from the Merrifield-Kuijken 
broadening function (Eq. (\ref{mkbf}).
For each value of S/N, 300 noise realizations of the same spectrum
were generated and the value of $\alpha$ that minimized the average
square deviation between the true and estimated $N(V)$'s was 
found.
The ordinate is the MISE of the estimates using this optimal $\alpha$.

\figcaption[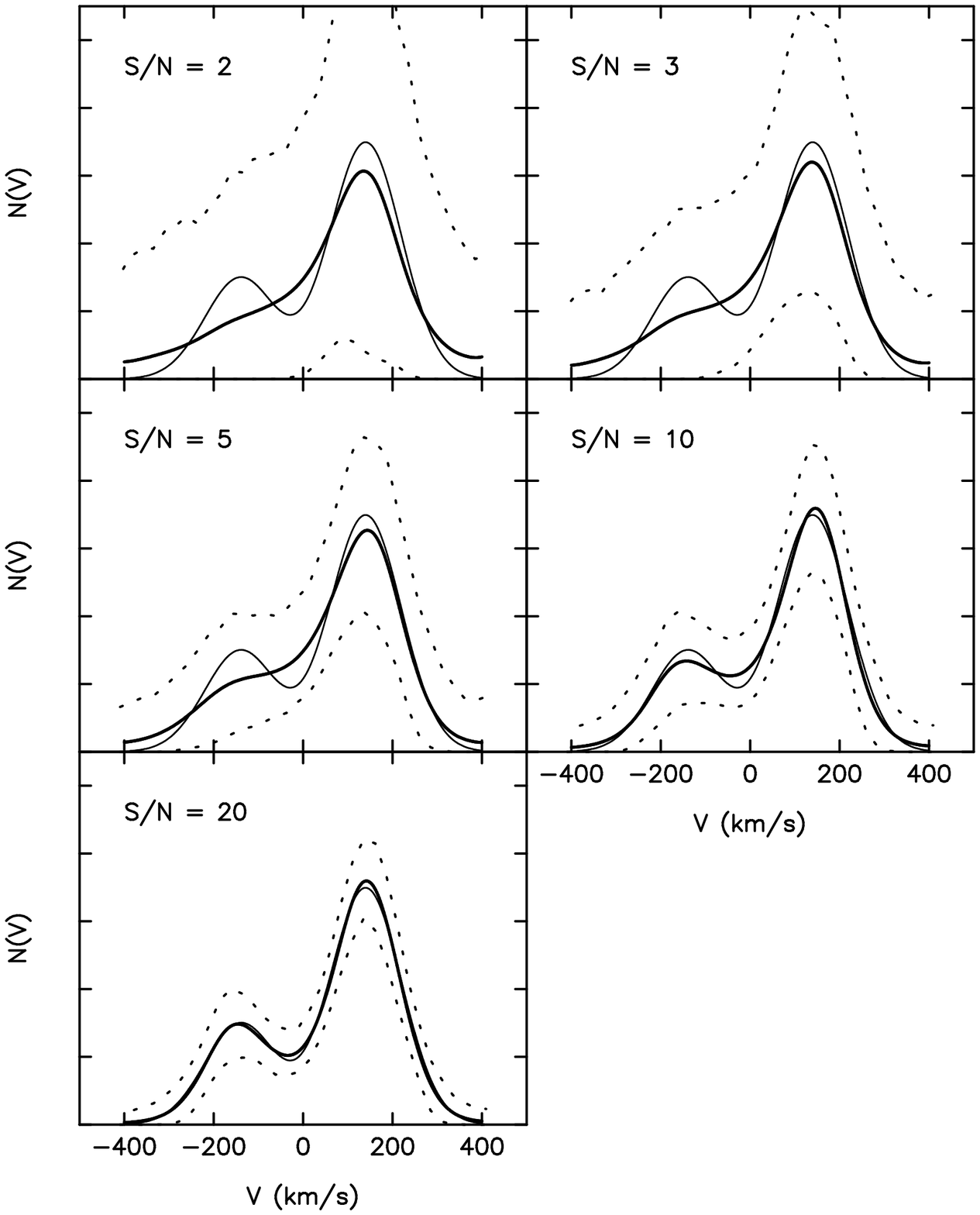]{\label{fig3}} Average estimates of $N(V)$, and 
their 95\% variance bands, based on 300 noise realizations of 
spectra generated from the Merrifield-Kuijken broadening 
function (\ref{mkbf}).
As the signal-to-noise ratio increases, the average estimate tends
toward the true broadening function and the variance of the
estimates decreases.

\figcaption[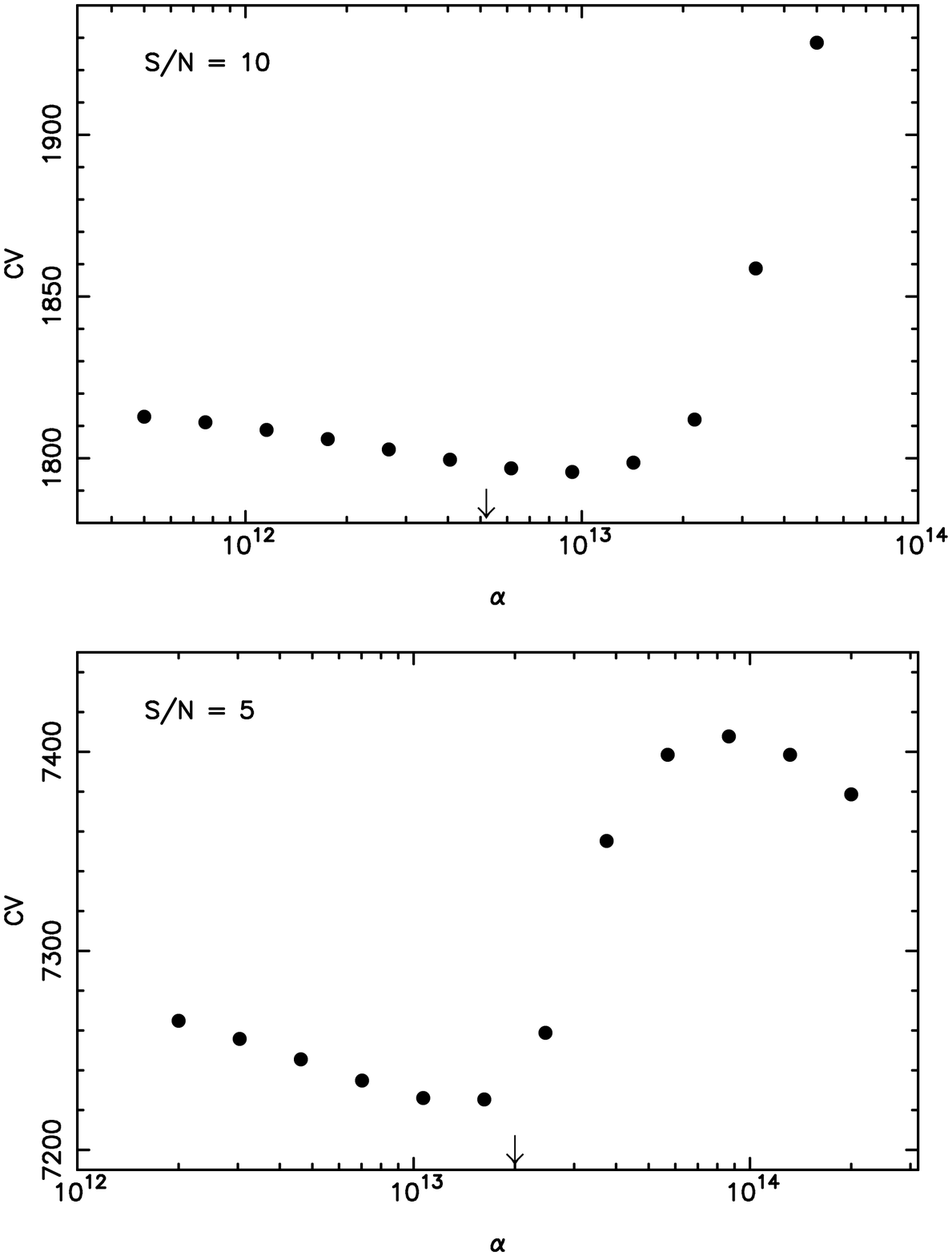]{\label{fig4}} Dependence of the
cross-validation score (CV) on the smoothing parameter $\alpha$ for two
spectra generated using the Merrifield-Kuijken broadening 
function (\ref{mkbf}), with ${\rm S/N} =10$ and ${\rm S/N} =5$.  
The value of $\alpha$ at the minimum of the CV curve 
is an estimate of the value of the smoothing
parameter that minimizes the integrated square error of
the estimated $N(V)$.
Arrows indicate the values of $\alpha$ that actually minimize the 
ISE of the broadening functions inferred from these two spectra.

\figcaption[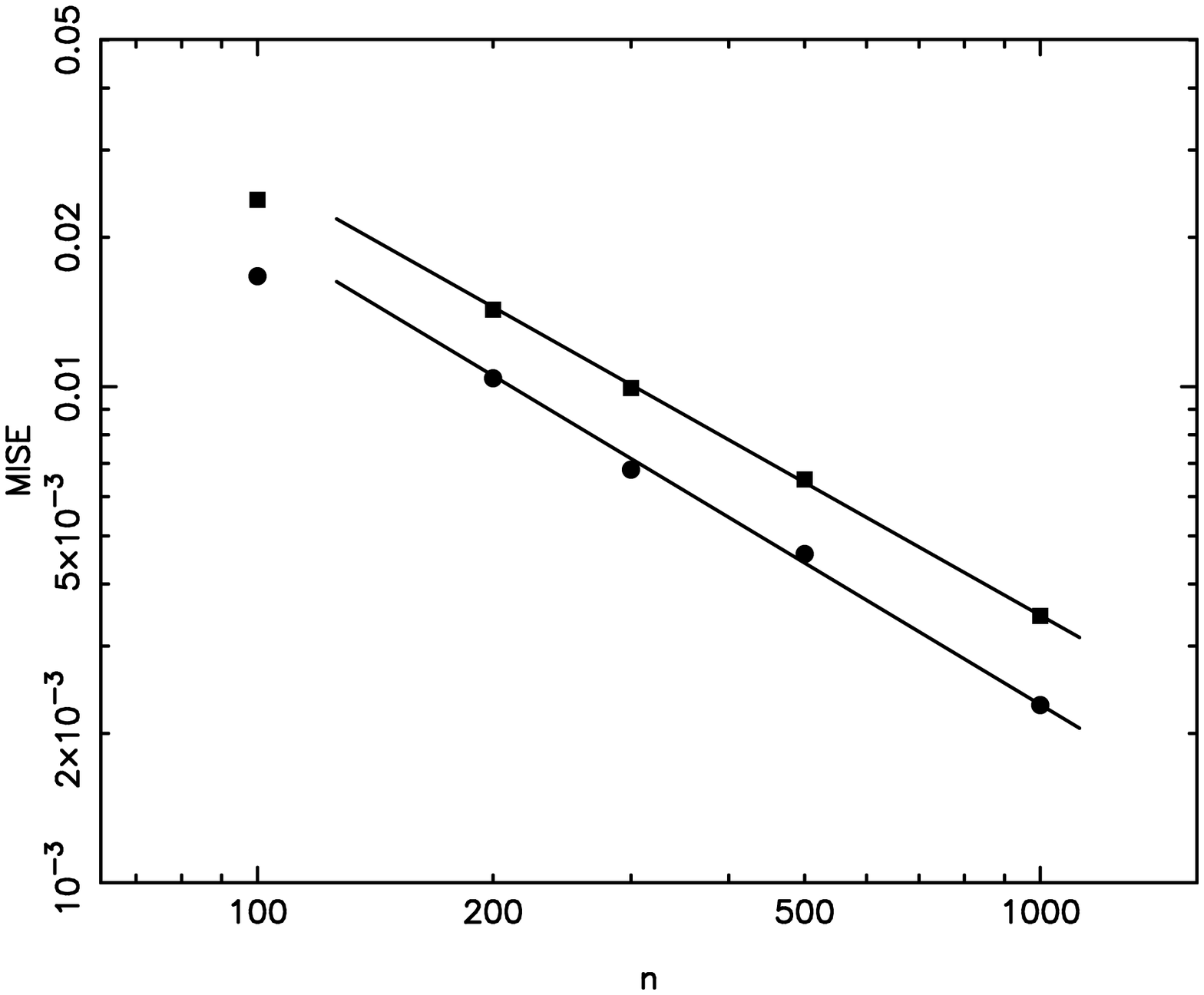]{\label{fig5}} Dependence of the 
mean integrated square error of the recovered $N(V)$ on the 
number of velocities $n$, for pseudo-data generated from the 
flat-topped velocity distribution of Eq. (\ref{flat}).
Squares: $\sigma_N=4$ km s$^{-1}$; circles: $\sigma_N=0$.

\figcaption[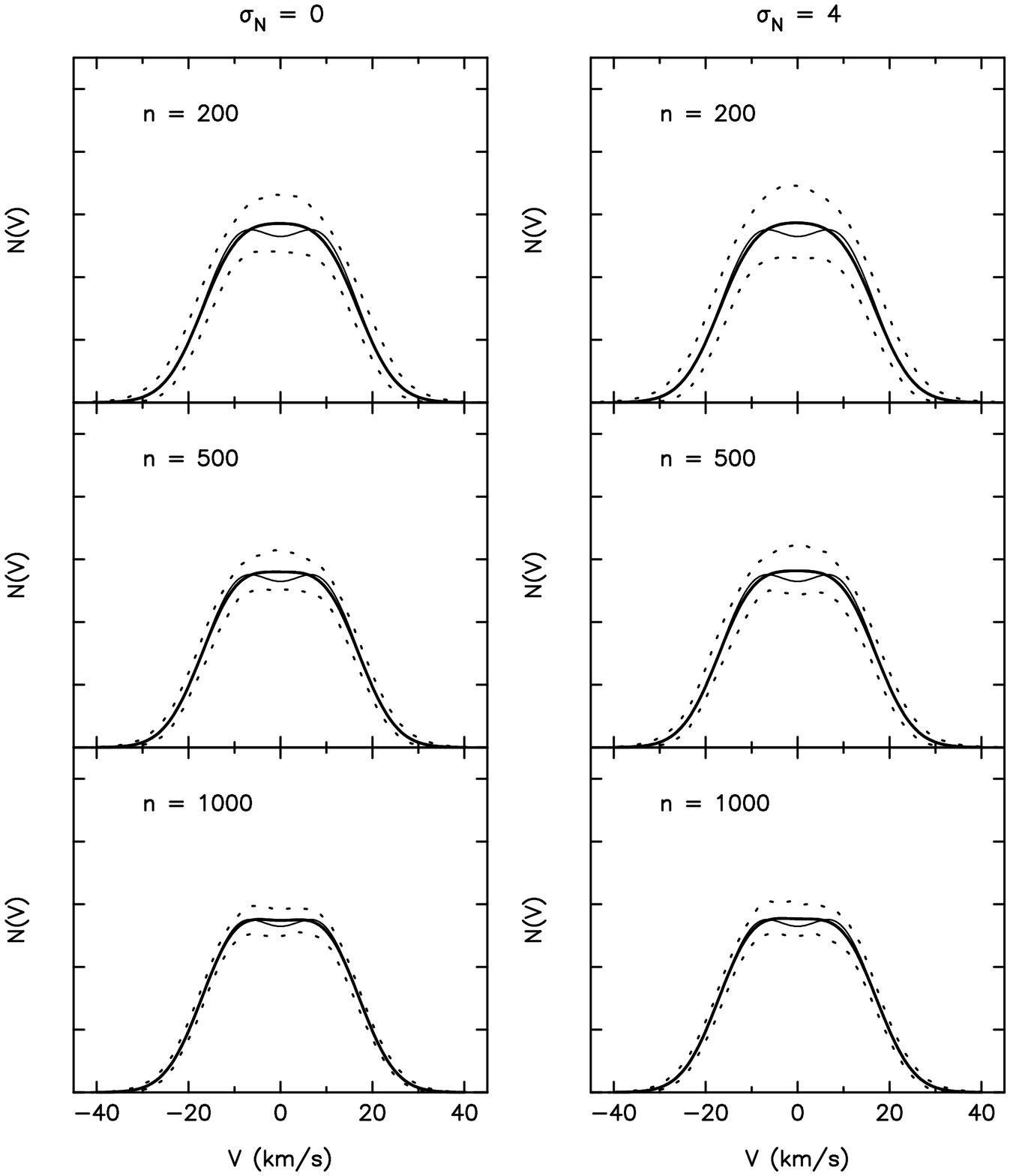]{\label{fig6}} Average estimates of $N(V)$, and 
their 95\% variance bands, based on 300 samples of discrete velocities 
generated from the flat-topped velocity distribution of Eq. (\ref{flat}).

\figcaption[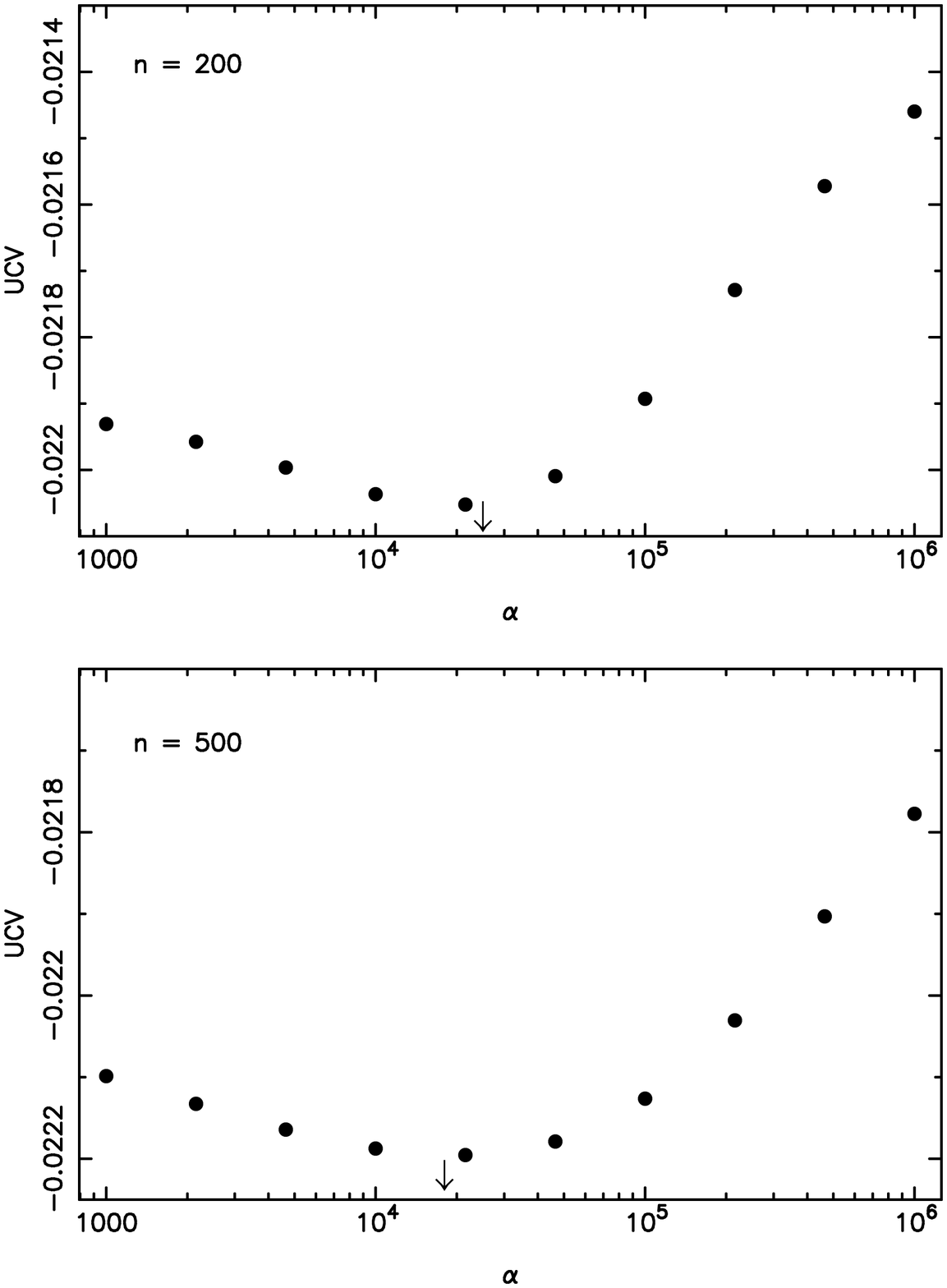]{\label{fig7}} Dependence of the unbiased
cross validation score (UCV) on the smoothing parameter $\alpha$ for 
data generated from the velocity distribution of Eq. 
(\ref{flat}), with $\sigma_N=0$.
The value of $\alpha$ at the minimum of the UCV curve 
is an estimate of the value that minimizes the integrated square error of
$\hat{N}(V)$.
Arrows indicate the values of $\alpha$ that actually minimize the 
ISE of the velocity distributions inferred from these two data sets.

\figcaption[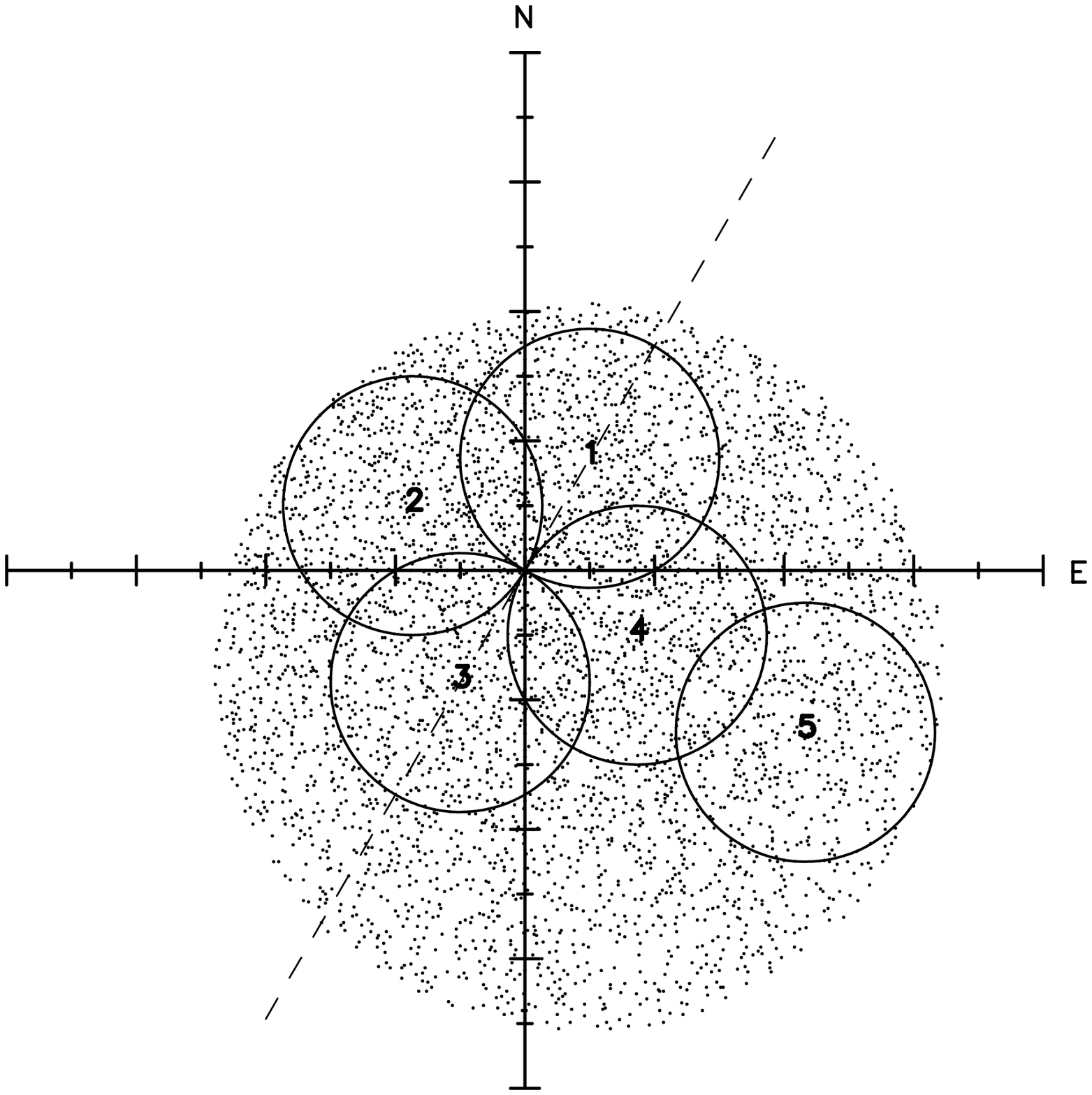]{\label{fig8}} Map of stars with observed
radial velocities in the globular cluster $\omega$ Centauri.
The origin of the coordinate system is the center of the cluster
as defined by Meylan \& Mayor (1986); major tick marks are separated
by one arc minute.
The Fabry-Perot field of view is offset by approximately 1.1 arc minute
from the origin.
The dashed line is an estimate of the projected rotation axis of the cluster.
Velocity distributions were computed for stars in the five circular
fields shown.

\figcaption[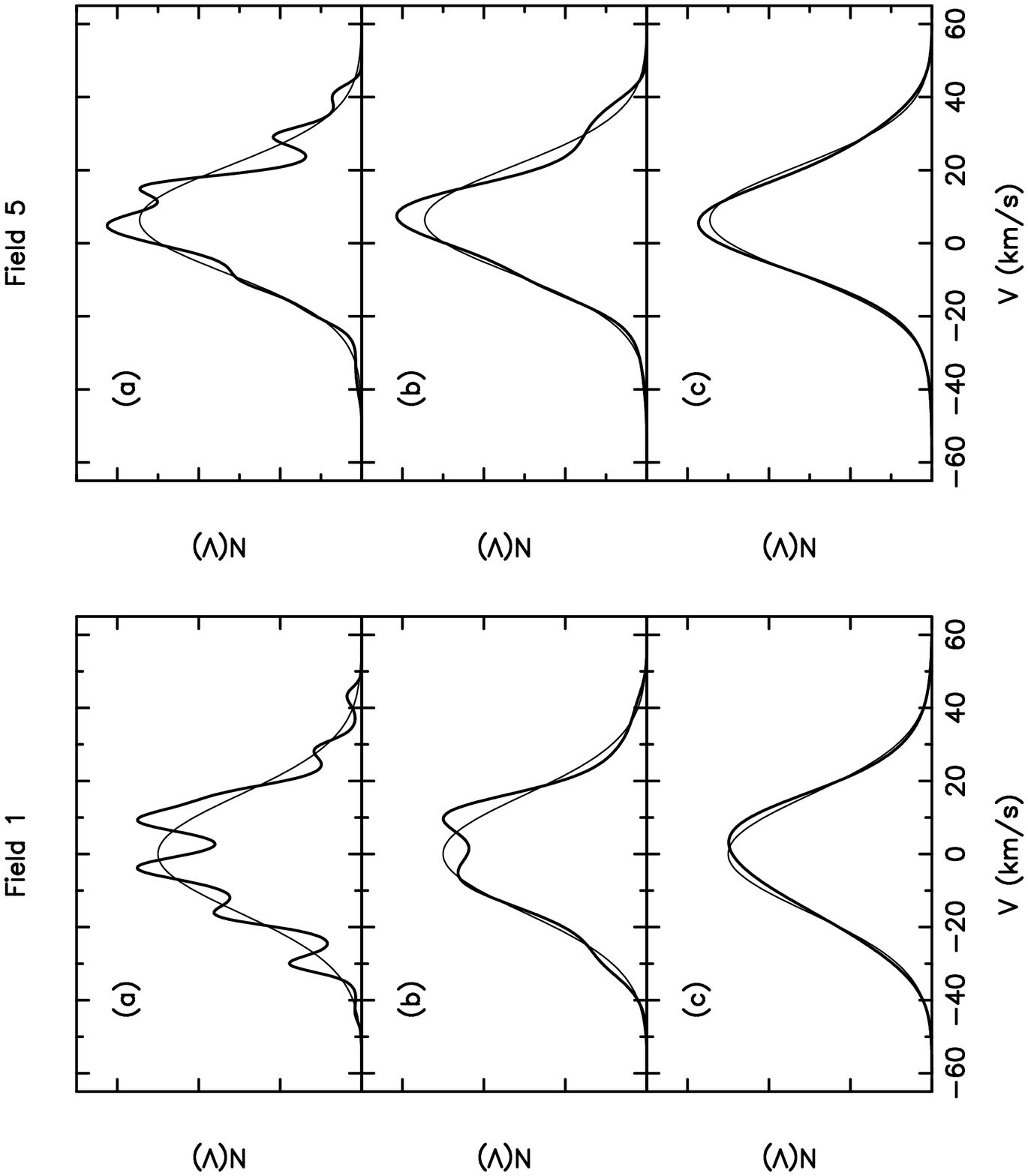]{\label{fig9}} Penalized-likelihood 
estimates of $N(V)$ in two fields in $\omega$ Centauri.
The value of the smoothing parameter $\alpha$ increases from (a) to (c).
Thin lines are normal $N(V)$'s with the same mean velocity and 
velocity dispersion as the estimated $N(V)$'s.

\figcaption[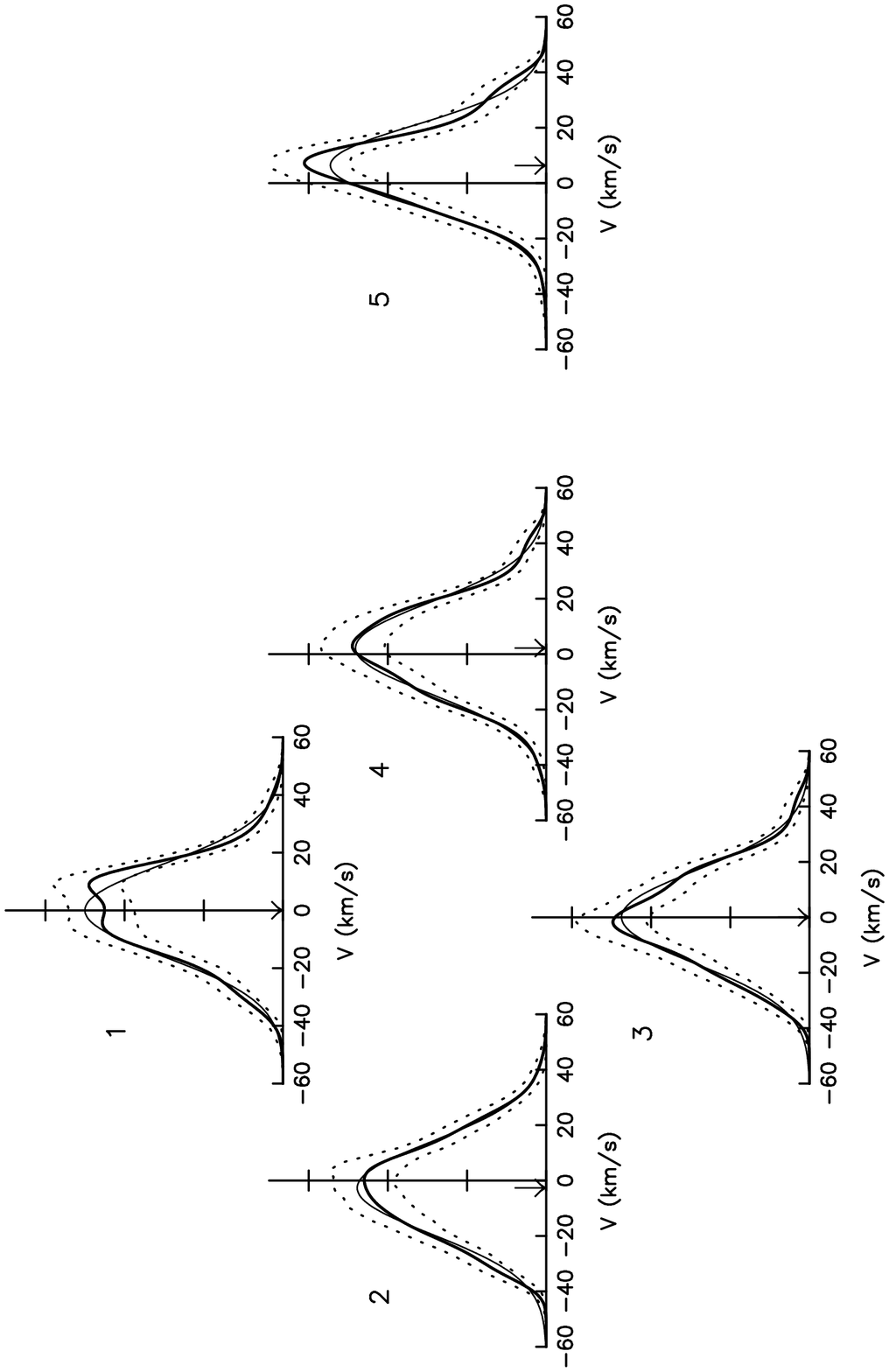]{\label{fig10}} Penalized-likelihood 
estimates of $N(V)$ in five fields in $\omega$ Centauri.
Heavy solid lines are the estimates; 
dashed lines are 95\% bootstrap confidence bands; thin solid lines are 
the normal distributions with the same mean velocity and velocity 
dispersion as the estimated $N(V)$'s.
Arrows indicate mean velocities.

\clearpage

\plotone{figure1.ps}

\clearpage

\plotone{figure2.ps}

\clearpage

\plotone{figure3.ps}

\clearpage

\plotone{figure4.ps}

\clearpage

\plotone{figure5.ps}

\clearpage

\plotone{figure6.ps}

\clearpage

\plotone{figure7.ps}

\clearpage

\plotone{figure8.ps}

\clearpage

\plotone{figure9.ps}

\clearpage

\plotone{figure10.ps}

\end{document}